\documentclass[10pt,onecolumn,preprintnumbers,amsmath,amssymb]{article}

\usepackage{geometry}
\usepackage{amsmath}  

\usepackage{graphicx}
\usepackage{dcolumn}
\usepackage{bm}
\usepackage{hyperref}
\usepackage{url}
\usepackage{epstopdf}
\usepackage{ragged2e}
\usepackage{abstract} 
\usepackage{titlesec} 

\titleformat{\section}[block]{\large\scshape\centering}{\thesection.}{1em}{} 
\titleformat{\subsection}[block]{\large}{\thesubsection.}{1em}{} 

\begin{document}

\title{\vspace{-1.5cm}\textbf{Tonal consonance parameters link microscopic and macroscopic properties of music exposing a hidden order in melody}}
\date{}
\maketitle
\centering{
	\author{J. E. Useche \footnote{\href{mailto:jeusecher@unal.edu.co}{jeusecher@unal.edu.co}}}
\author{R. G. Hurtado\footnote{Corresponding author. Postal address: Carrera 30 No. 45-03, Departamento de F\'{\i}sica, Universidad Nacional de Colombia, Bogot\'a, Colombia. \href{mailto:rghurtadoh@unal.edu.co}{rghurtadoh@unal.edu.co}}}
\\
}%
 


\begin{abstract}
	\noindent Consonance is related to the perception of pleasantness arising from a combination of sounds and has been approached quantitatively using mathematical relations, physics, information theory, and psychoacoustics. Tonal consonance is present in timbre, musical tuning, harmony, and melody, and it is used for conveying sensations, perceptions, and emotions in music. It involves the physical properties of sound waves and is used to study melody and harmony through musical intervals and chords. From the perspective of complexity, the macroscopic properties of a system with many parts frequently rely on the statistical properties of its constituent elements. Here we show how the tonal consonance parameters for complex tones can be used to study complexity in music. We apply this formalism to melody, showing that melodic lines in musical pieces can be described in terms of the physical properties of melodic intervals and the existence of an entropy extremalization principle subject to psychoacoustic macroscopic constraints with musical meaning. This result connects the human perception of consonance with the complexity of human creativity in music through the physical properties of the musical stimulus.
\end{abstract}


\section{Introduction}
\justify
Pythagoras found that two sounds emitted simultaneously by vibrating strings of equal tension and density produce a pleasant sensation when the ratio between their lengths (and hence their fundamental frequencies) corresponds to the ratio between two small natural numbers ${n}/{m}$ \cite{Rossing,Roderer,Regnault}. This sensation is formally defined as consonance, and it is present in melody, harmony, timbre, and musical tuning \cite{Roderer,Regnault,SetharesArticle}. Many authors relate consonance to conveying musical information as emotions and meaning \cite{Copland,Madell,Schopenhauer,Ball}. From the perspective of the nature of tonality, consonance and dissonance give rise to emotions through tension and relaxation \cite{Madell} in passages from satisfaction to dissatisfaction and again to satisfaction \cite{Schopenhauer,Budd}.
\\ A starting point for studying consonance in psychoacoustics is the superposition
of pure tones with different frequencies \cite{Rossing,Helmhotz,Heffernan}. Hermann von Helmholtz found that the consonance level of pairs of simultaneous pure tones is related to the beats produced by fluctuations in the peak intensity of the resulting sound waves \cite{Helmhotz}; specifically, the perception of dissonance is proportional to the perception of roughness due to rapid beats \cite{Regnault}.
\\Musical instruments produce complex tones that can be represented by the superposition of several pure tones. The corresponding set of frequencies with their amplitudes is called the spectrum which strongly characterizes the timbre of a musical instrument \cite{Rossing}. For complex tones, Helmholtz inferred that the superposition of spectrum components with close frequencies is related to the perception of dissonance  \cite{Helmhotz,Regnault}.
\\After Helmholtz, Reinier Plomp and Willem Levelt reported that the transition range between consonance and dissonance is related to a critical bandwidth that depends on the frequency difference of the corresponding sound waves \cite{Plomp}. This approach to consonance is known as tonal or sensory, because it depends on the physical properties of the stimulus, regardless of the cultural context of the listener \cite{Schellenberg}.
\\William Sethares assigned a level of tonal consonance to timbre using the spectrum of the emitted sound and connected timbre with musical tuning \cite{SetharesArticle,SetharesBook}. 
Musical tuning refers to adjusting a set of pitches to a musical scale using a fixed pitch of reference. Pitch is a subjective quality of sound that judges its height and depends strongly on the lowest frequency of the spectrum, the fundamental frequency \cite{Rossing}, and usually a musical scale is a set of mathematical relations among the fundamental frequencies of pitches. Pairs of pitches in a musical scale define musical intervals of size $ L $ given by the number of pitches between them. In musical theory, the level of consonance assigned to a musical interval usually depends on its size (and hence the fundamental frequency ratio of its pitches) \cite{Aldwell}. Since musicians tend to apply the same rules for judging the consonance level of simultaneous and successive pitches (harmonic and melodic intervals respectively), and the short-term persistence of pitch in auditoriums may give rise to consonance sensations for successive pitches \cite{Regnault}, then tonal consonance is suitable for analyzing both harmony and melody.
\\In this paper we will study the consonance properties of melody, defined by the New Grove Dictionary of Music and Musicians as ``pitched sounds arranged in musical time in accordance with given cultural conventions and constraints'' \cite{Apel}. An alternative definition that encompasses music and speech was given by Aniruddh Patel: ``an organized sequence of pitches that conveys a rich variety of information to a listener'' \cite{Patel}.
\\With respect to the statistical properties of melody, George Kingsley Zipf studied the frequency of occurrence of melodic intervals in masterpieces of Western tonal music. Melodic intervals can be played in an ascendant or a descendant manner, and Zipf reported that the frequency of occurrence in both cases is almost inversely proportional to their size \cite{Zipf}. Melodies tend to meander around a central pitch range, and for many cultures an asymmetry emerges in this meandering, in the sense that large melodic intervals are more likely to ascend and small melodic intervals are more likely to descend \cite{Huron}. A more recent study proposes long-tailed Levy-stable distributions to model the probability distribution of melodic intervals as a function of their size \cite{Niklasson}. The use of physical quantities to represent melodic intervals has generated a new approach to analyze melody obtaining exponential and power law probability distributions with good experimental determination coefficients \cite{CogSci}. These studies show signs of complexity in melody but don't define a formal relation between the size of musical intervals and consonance. Other studies also show that musical pieces present complexity as scale-free patterns in the fluctuations of loudness \cite{Voss}, rhythm synchronization \cite{Hennig}, and pitch behavior in melody \cite{Voss}, as well as in the connectivity properties of complex networks representing successive notes of musical pieces \cite{Liu}. Regarding consonance, a recent study found scale-free patterns in the consonance fluctuations associated with harmony \cite{Wu}.
\\From the perspective of information theory, G\"ung\"or G\"und\"uz and Ufuk G\"und\"uz measured the probability of occurrence of musical notes during the progress of melodies and found that entropy grows up to a limiting value smaller than the entropy of a random melody \cite{Gunduz}.
\\In this paper, we show how tonal consonance parameters for complex tones can be used to study complexity in music, using a more general quantity than the size of the musical interval, because it distinguishes the level of consonance and the position in the register of a musical interval. In order to demonstrate the usefulness of this formalism, we apply it to melody through the study of twenty melodic lines from seven masterpieces of Western tonal music. We also develop a theoretical model based on relative entropy extremalization, in agreement with the qualitative definitions of melody, for reproducing the main features of experimental results and linking the microscopic tonal consonance properties of melodic intervals, including timbre, with the macroscopic ones stemming from their organization in real melodic lines.
 
\section{Tonal consonance parameters for pure and complex tones}
The method for determining the tonal consonance of pure tones is related to the beats produced by the superposition of two sinusoidal signals with different frequencies, $f_{i}$ and $f_{j}$. The superposition of pure tones varies in time with a rapid frequency $(f_{i}+f_{j})/2$ modulated by a slow one $|f_{j}-f_{i}|/2$.
The beats produced by fluctuations in the peak intensity of sound waves occur with a frequency $|f_{j}-f_{i}|$, and this phenomenon is independent of the differences in amplitude and phase between the two pure tones (see  supplementary material) for details). This result indicates that the main contribution of tonal consonance for pure tones are their frequency components. 
\\The approach of Plomp and Levelt to tonal consonance of complex tones is independent of musical scales, and they found that an interval of a given size $ L $ might be more or less consonant depending on its timbre and location in the register and that this variation through the register is continuous and smooth \cite{Plomp}. They used two quantities for parametrizing the tonal consonance level of complex tones: the lowest fundamental frequency of the pair of pitches and the ratio between the fundamental frequencies $f_{j}/f_{i}$ \cite{Plomp}. This set of parameters is equivalent to one with the same ratio $f_{j}/f_{i}$  and the absolute value of the difference between the fundamental frequencies $ |f_{j}-f_{i}|$ (see supplementary material for details), which will be the equivalent parameter in comparison with pure tones. We use the William Sethares formalism to reproduced these curves in the case of a particular timbre (see supplementary material for details). 
\\Figure \ref{ConsonanceCurve}a shows the tonal consonance curves in the case of simultaneous pitches (which for musicians can also be used in the case of successive pitches) for a timbre of six harmonics of the same amplitude as the fundamental, in the case of the frequency ratios of the twelve-tone equal-tempered scale within an  octave. In this figure, we can see that an interval of a certain size  $ L $ is more consonant in the middle part of the registry than in the lowest part. For all sizes, these curves fit to a second-order exponential decay function with a coefficient of determination $R^{2}=0.99$. 
\begin{figure}[htb]
	\centering
	\begin{tabular}{c}
		\includegraphics[width=8.0cm]{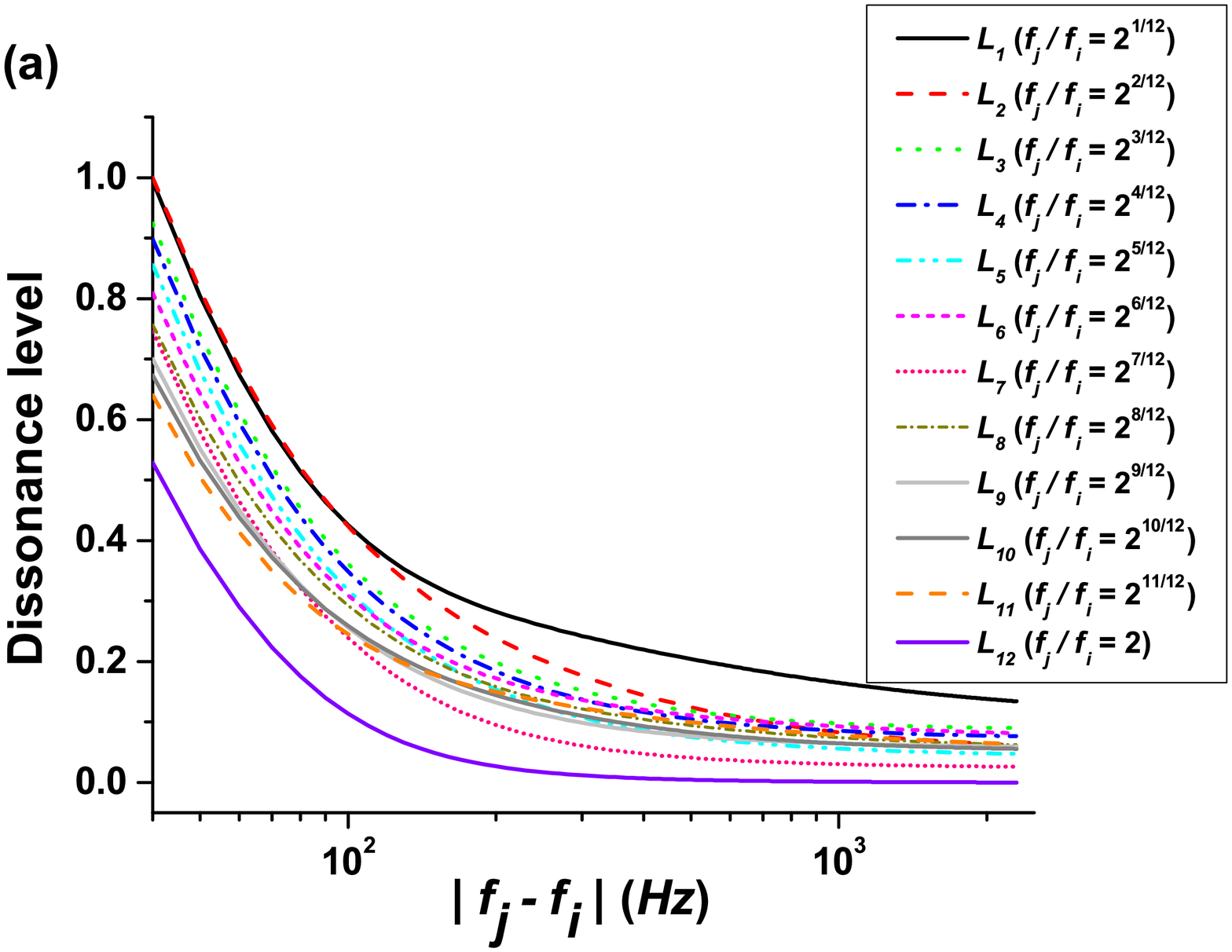}
	\includegraphics[width=8.0cm]{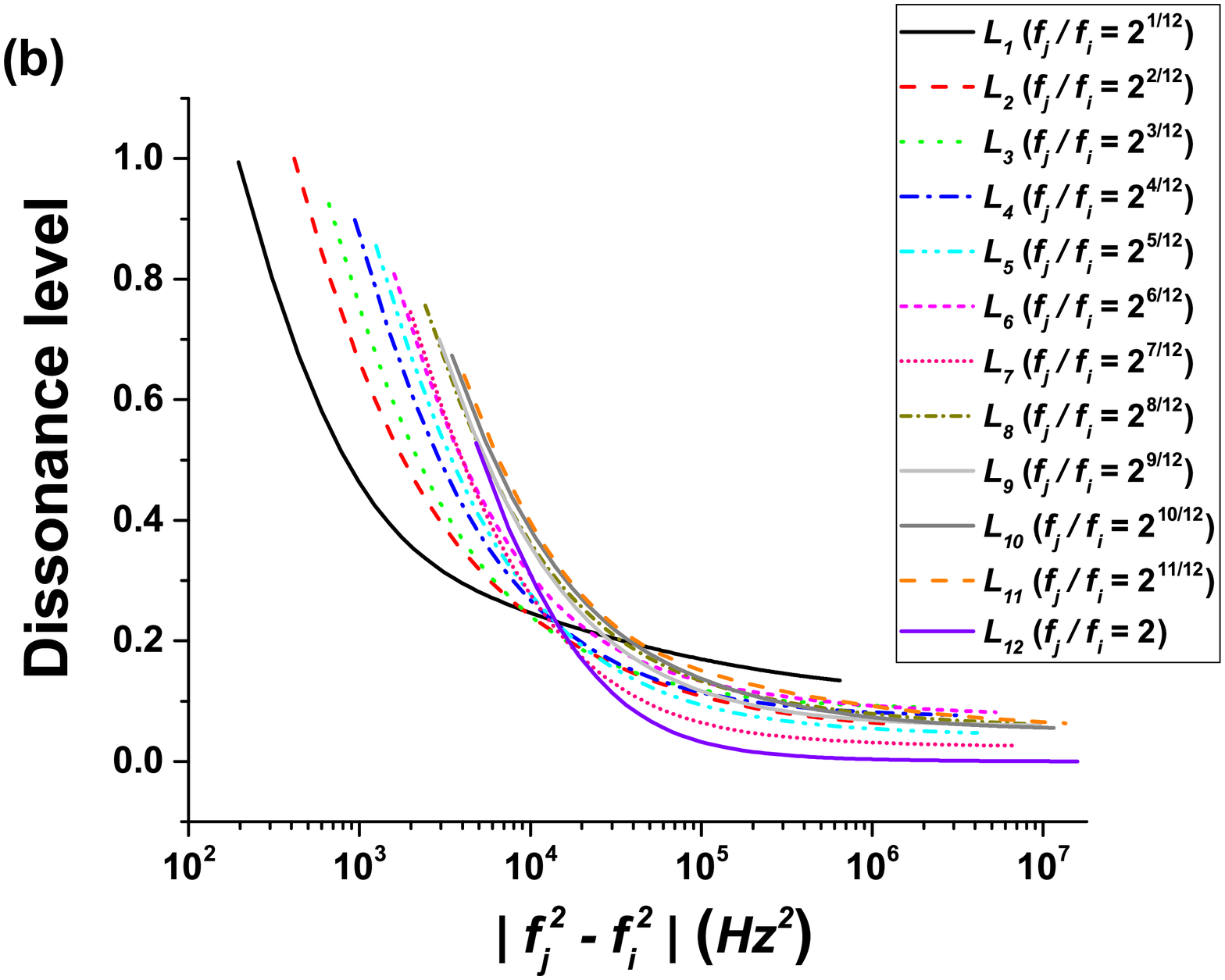}
	\end{tabular}
	\caption{(a) Relation between the dissonance level and the absolute value of the frequency difference. (b) Relation between the dissonance level and the absolute value of the difference in the squares of the frequencies. In both cases we consider a timbre of six harmonics of the same amplitude as the fundamental. Each possible size $ L $ corresponds to a particular frequency ratio inside the octave in the twelve-tone equal-tempered scale. The dissonance level scale has been normalized between $ 0 $ and $ 1 $ for the typical register of an eighty-eight key piano.}\label{ConsonanceCurve}  
\end{figure} 

\section{Representation of a pair of pitches in musical scales using tonal consonance parameters}
\begin{figure}[htb]
	\centering
	\begin{tabular}{c}
		\includegraphics[width=8.0cm]{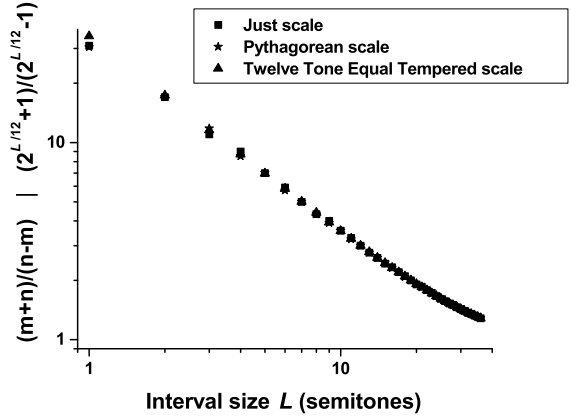} 
	\end{tabular}
	\caption{Relation between musical scale parameters and the interval size for the just, Pythagorean, and twelve-tone equal-tempered scales. Interval size from one to thirty-six semitones. $ n>m $ and $ L>0 $.}\label{JustaPitTemp}  
\end{figure}

\begin{figure}[htb]
	\centering
	\begin{tabular}{c}
		\includegraphics[width=8.0cm]{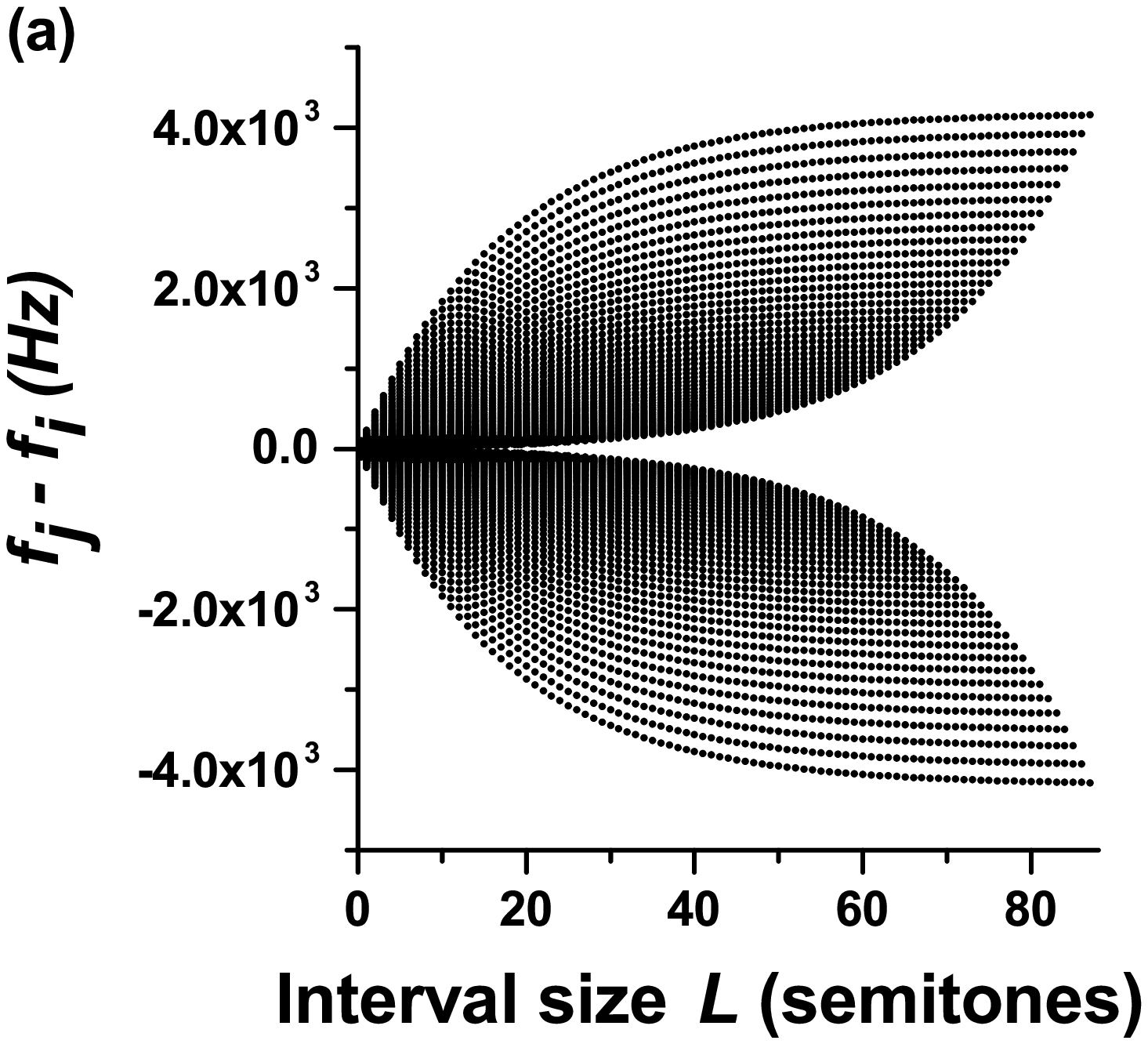}
		\includegraphics[width=8.0cm]{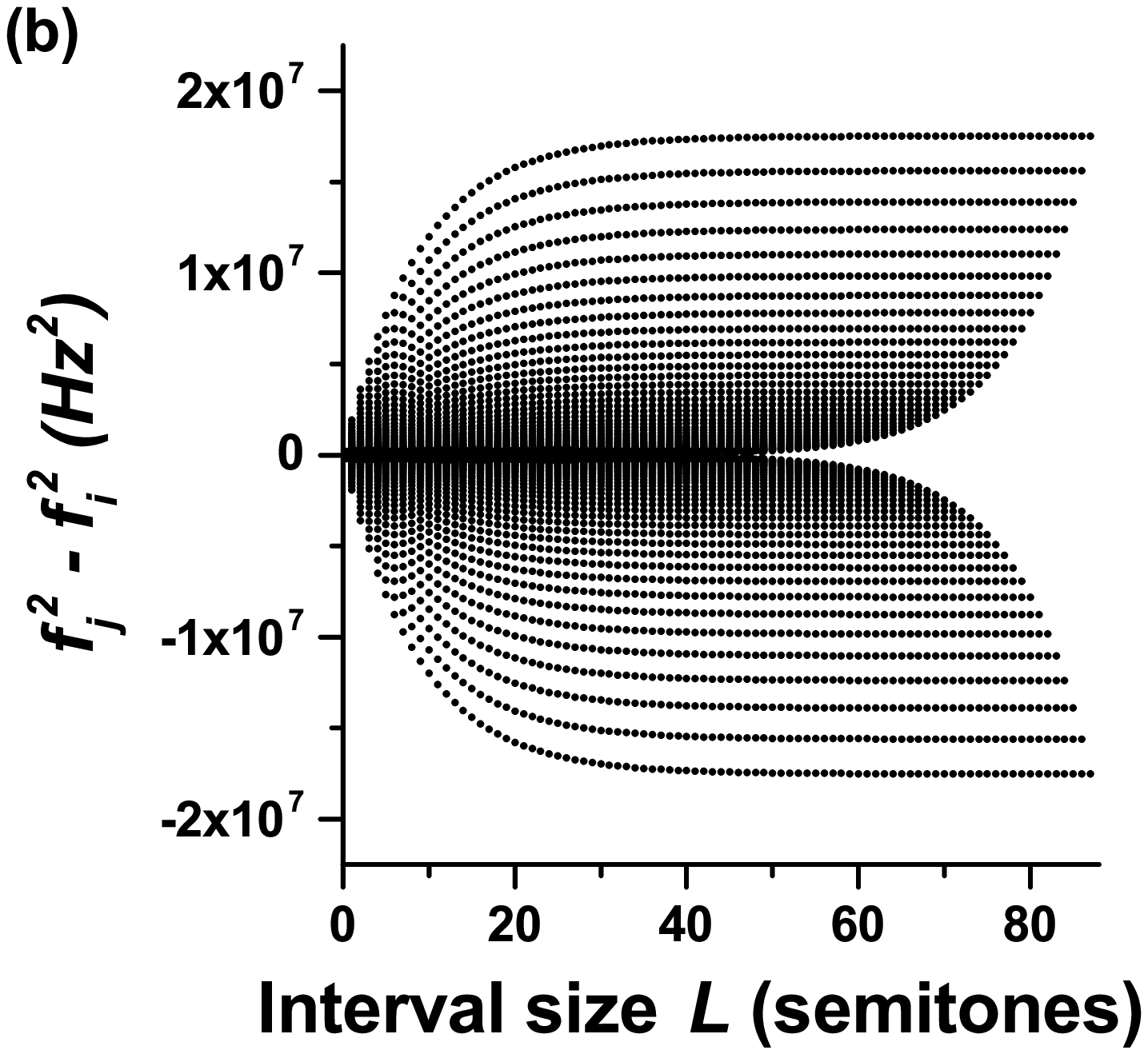}
	\end{tabular}
	\caption{Relation between the quantities $ f_{j}-f_{i} $ and $ f_{j}^{2}-f_{i}^{2} $ and the interval size in semitones ((a) and (b) respectively) for the register of a typical eighty-eight key piano. These quantities distinguish intervals of equal size played in different parts of the register and between intervals of different sizes. The upper branch comes from $ j=88 $ (highest pitch) and $ i $ varies from $ 88 $ to $ 1 $. The tuning comes from the frequency relation for the twelve-tone equal-tempered scale with $A=440\:Hz $.}\label{Distinction}  
\end{figure}

For scales based on the Pythagorean rule, as for example the just and the Pythagorean \cite{Rossing,Roderer}, the difference of fundamental frequencies is related to their sum through
\begin{equation}\label{SumDifference}
f_{j}-f_{i}=[(n-m)/(n+m)](f_{j}+f_{i})\quad.	
\end{equation}
For the just and the Pythagorean scales, the quantity $ (n-m)/(n+m) $ depends on the size $ L\equiv L(f_{i},f_{j}) $ of the corresponding interval (see Figure \ref*{JustaPitTemp}), and for $ f_{j}\neq f_{i} $ (i.e. $ L\neq 0 $) this relation can be expressed as: 
\begin{equation}\label{SumDifferenceL}
f_{j}+f_{i}=(a)(-1)^{h}L^{-b}(f_{j}-f_{i})\quad, 
\end{equation}	
with $ h=0 $ for $ f_{j}>f_{i} $ and $ h=1 $ for $ f_{j}<f_{i} $. Up to three octaves, since melodic intervals in musical pieces usually do not exceed this size \cite{Aldwell,Patel}, the fitting parameters to a power law for the Pythagorean scale are $ a=30.801\pm 0.184 $ and $ b=0.918 \pm 0.006 $, with $ R^{2}=0.9988 $, and for the just scale are $ a=31.176\pm0.149 $ and $ b=0.925\pm0.005 $, with $ R^{2}=0.9992 $. The frequency ratios used to construct these scales  are presented in supplementary material in agreement with \cite{Rossing}.
Equation \ref{SumDifference} does not hold for tempered scales, however, since pitches in the twelve-tone equal-tempered scale are given by $ f_{i}=f_{1}\sqrt[12]{2^{i}}$, where $ f_{1} $ is a reference frequency. Then
\begin{equation}\label{RatioTemp}
f_{j}/f_{i}=\sqrt[12]{2^{j-i}}=\sqrt[12]{2^{L}},\:  \textnormal{with}\: f_{j}>f_{i}\quad, 		
\end{equation}
and for $ f_{j} \neq f_{i}  $ : 
\begin{equation}\label{SumDifferenceTemp}
f_{j}+f_{i}=\dfrac{2^{L/12}+1}{2^{L/12}-1}(f_{j}-f_{i})\quad .
\end{equation}
In this case, the fit parameters in Equation \ref{SumDifferenceL} are $ a = 34.456\pm 0.139 $ and $ b = 0.979\pm 0.004 $, with $ R^{2}  = 0.9994 $, see Figure \ref{JustaPitTemp}.
\\Hence for these scales, the sum of the fundamental frequencies contains information about the height of pitches, and from Equations \ref{SumDifferenceL} and \ref{SumDifferenceTemp}, about the tonal consonance parameter $ |f_{j}-f_{i}| $ per unit of interval size $L$, as $ b\approx 1 $.
\\Since each ratio $f_{j}/f_{i}$ corresponds to a size $ L $ and it depends on the ratio $ (f_{j}-f_{i})/(f_{j}+f_{i}) $, then a complete description of tonal consonance can be made using the sum and the difference of the fundamental frequencies. Additionally, the set of these two quantities distinguishes each pair of pitches, since there are two equations with two variables, therefore relating the use of tonal consonance in a musical piece with the selection of pitches made by the composer. Constructing probability distributions in musical pieces over these two quantities, it is possible to infer the use of tonal consonance associated with harmonic or melodic intervals. We propose to study only one probability distribution containing information about these two quantities in such a way that for each possible combination $[f_{j}+f_{i},f_{j}-f_{i}]$ it will be possible to associate only one possible value. A quantity that combines the two parameters selected for describing tonal consonance for complex tones is $f_{j}^{2}-f_{i}^{2}=(f_{i}+f_{j})(f_{j}-f_{i}) $, for scales with unique values of this quantity for each pair of pitches, such as the twelve-tone equal-tempered scale and surely in all cases of the just and the Pythagorean scales (see supplementary material for details). This quantity allows reconstructing both parameters, $ (f_{i}+f_{j}) $ and $(f_{j}-f_{i}) $, assigning a tonal consonance level to each pair of pitches. 
Figure \ref{ConsonanceCurve}b shows the dissonance level for the case of a timbre of six harmonics of the same amplitude as the fundamental, as a function of the quantity $f_{j}^{2}-f_{i}^{2}$ for each possible size $ L $, in the twelve-tone equal-tempered scale within the octave. This figure can be produced using the William Sethares Method (see supplementary material) or alternatively using the second-order exponential decay functions of Figure \ref{ConsonanceCurve}a, which relates consonance with the absolute value of the frequency difference and the relation for musical scales:
\begin{equation}
f_{j}^{2}-f_{i}^{2}=(f_{i}+f_{j})(f_{j}-f_{i})=(a)(-1)^{h}L^{-b}{|f_{j}-f_{i}|}^2. 
\end{equation}
A curious property of this quantity is that if two sound waves propagate in the same medium (with density $ \rho $) and we are in a musical domain in which it is possible to assume equal amplitudes $ T $, then the quantity $ f_{j}^2-f_{i}^2$ is proportional to the difference of the average density of the total energy carried by the two waves \cite{Pain}:
\begin{equation}\label{Energy}
\epsilon_{j}-\epsilon_{i}=2\pi^{2}\rho T (f_{j}^{2}-f_{i}^{2})\;. 
\end{equation}
This relation holds for pure tones and, in the case of complex tones, it corresponds to the difference of the average of the energy density carried by the fundamental components. 
\\We found that the quantity $ f_{j}-f_{i} $ also produces different values for the studied scales, so we also present the probability distributions of this quantity. As we explain in the next section, the quantity $ f_{j} + f_{i} $ is  useless for our analysis because it can't distinguish between the chronological order of the pitches, between ascending and descending intervals.
\\The magnitudes of $ f_{j}-f_{i} $ and $ f_{j}^{2}-f_{i}^{2} $ distinguish intervals of equal size played in different parts of the register and between intervals of different size, except for unisons $ f_{j}=f_{i} $ with a degenerated value of $ 0 $; see Figure \ref*{Distinction}. 
\\For the case of $ f_{j}^{2}-f_{i}^{2} $ the distinction between intervals is better between one and approximately thirty-six semitones. However, in this region (the musically relevant one \cite{Patel}), the relative distances between different values are larger in the case of the quantity $ f_{j}^{2}-f_{i}^{2} $ than in the case of the quantity $ f_{j}-f_{i} $. This behavior can be observed in Figure \ref*{Distinction} through  the order of magnitude of the values and the relative separation between branches.    
\section{Analysis of real melodic lines}
\begin{figure}[htb]
	\centering
	\begin{tabular}{c}
		\includegraphics[width=8.0cm]{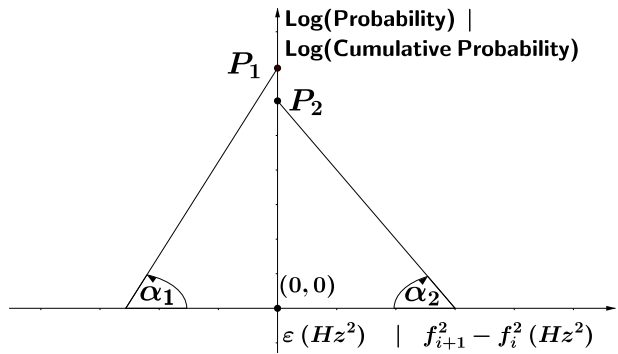}\end{tabular}
	\caption{General form of the probability distribution $ P(\varepsilon) $ and cumulative probability distribution $ P(f_{i+1}^2-f_{i}^2) $. In the symmetric case, $ P_{1}=P_{2} $ and $ \alpha_{2}=\alpha_{2} $.}\label{AsymmetricLaplace}  
\end{figure}

\begin{figure}[htb]
	\centering
	\begin{tabular}{c}
		\includegraphics[width=8.0cm]{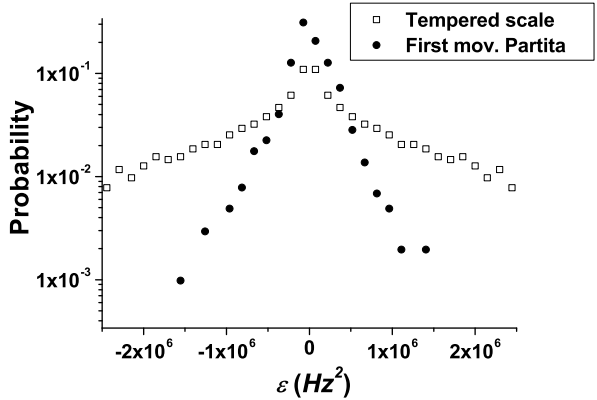}\end{tabular}
	\caption{Twelve-tone equal-tempered scale effect. Comparison between the probability distribution for the real melodic line of the first movement of the \textit{Partita in A minor BWV 1013} by J. S. Bach and the corresponding bin degeneration for the same \textit{ambitus}.}\label{TempScalePartita}  
\end{figure}

We study the use of tonal consonance regarding the selection made by the composer of melodic intervals characterized by their size and position in the register. For this purpose, we study the probability distribution of the physical quantities $ f_{j}-f_{i} $ and $ f_{j}^{2}-f_{i}^{2} $. 
\\If $ i $ indicates the chronological order of appearance of pitches in a melody, then the quantities $ f_{i+1}-f_{i} $ and $ f_{i+1}^{2}-f_{i}^{2} $ can be used to study tonal consonance with the sign distinguishing between ascending ($ f_{i+1}>f_{i} $) and descending ($ f_{i+1}<f_{i} $) transitions (see Figure \ref{Distinction}). We analyze vocal and instrumental pieces of the Baroque and Classical periods played in the twelve-tone equal-tempered scale with $ A=440\:Hz $. The selected pieces contain melodic lines characterized by their great length, internal coherence, and rich variety of instruments and registers \cite{Aldwell}.\\
\textit{Brandenburg Concerto No. 3 in G Major BWV 1048}. Johann Sebastian Bach: The polyphonic material in this concerto for eleven musical instruments (three violins, three violas, three cellos, violone, and harpsichord) makes it possible to assume that each instrument has a melodic motion.\\
\textit{Missa Super Dixit Maria}. Hans Leo Hassler: Polyphonic composition for four voices (soprano, contralto, tenor and bass).\\
\textit{First movement of the Partita in A Minor BWV 1013}. Johann Sebastian Bach: This piece has just one melodic line, for flute.\\ 
\textit{Piccolo Concerto RV444}. Antonio Vivaldi (arrangement by Gustav Anderson): We selected the piccolo melodic line because of its rich melodic content.\\
\textit{Sonata KV 545}. Wolfgang Amadeus Mozart: We selected the melodic line for the right hand of this piano sonata assuming that it drives the melodic content.\\
\textit{Suite No. 1 in G Major BWV 1007} and \textit{Suite No. 2 in D Minor BWV 1008}. Johann Sebastian Bach: The melodic lines of these pieces written for cello have mainly successive pitches. In the case of the few simultaneous pitches, the continuation of the melodic lines was assumed in the direction of the highest pitch.\\
First, we generated simplified MIDI files \cite{Liu} in order to extract from the scores the probability distributions of the quantities $ f_{i+1}-f_{i} $ and $ f_{i+1}^{2}-f_{i}^{2} $ and the complementary cumulative distribution functions (CCDF) for $ |f_{i+1}-f_{i}| $ and $ |f_{i+1}^{2}-f_{i}^{2}| $, which give information about the functional form of the probability distributions. Since the concept of \textit{chroma} states that the consonance level of the unison is equivalent to the corresponding octave \cite{Roderer} and the octave can be an ascending ($f_{i+1} / f_{i} =2/1$) or a descending ($f_{i+1} / f_{i} =1/2$) transition, then for our analysis we consider the unison as an ascending transition as well as a descending one. This means that we begin the bin count considering the bin box $[0,x)$ for ascending transitions and the bin box $(y,0]$ for descending transitions, where $x$ and $y$ are bin widths generated by the Sturges criterion in histograms. In our experimental analysis we found that the contribution of unisons is important for ascending transitions as well as for descending ones. Furthermore, as we have different right hand limit and left hand limit when we approach to $0$, we can't take a bin around $0$ containing ascending and descending transitions. Additionally, if we try to distribute the unisons between the ascending and the descending part, it procedure reduces the determination coefficient $R^{2}$ in histograms \cite{UsecheMaster} because we would be modifying the right hand and the left hand limits.\\
In the cases of ascending transitions ($ f_{i+1}>f_{i} $), descending transitions ($ f_{i+1}<f_{i} $ using $ |f_{i+1}-f_{i}| $ and $ |f_{i+1}^{2}-f_{i}^{2}| $), and the joint set of them (using $ |f_{i+1}-f_{i}| $ and $ |f_{i+1}^{2}-f_{i}^{2}| $ for all transitions), the CCDF fit to exponential functions for all sets with average determination coefficient $ \overline{R^{2}}\approx 0.99 $ with an standard deviation $SD \approx 0.01 $ (see supplementary material for details). Since the number of transitions in the studied melodic lines is at maximum one order of magnitude larger than the total number of possible transitions between pairs of successive pitches in the same \textit{ambitus} (range between the lowest pitch and the highest one) and the number of possible transitions for any melodic line is finite independently of its length, then the probability distributions must be represented in histograms with bin width moderately dependent on the number of transitions. These conditions are satisfied by the Sturges criterion \cite{Scott}. This analysis with histograms is important due to the typical lengths of melodic lines in music. 
\\Histograms for both quantities fit to exponential functions with the highest $ \overline{R^{2}} $ for $ f_{i+1}^{2}-f_{i}^{2} $ with ascending and descending transitions taken separately: for ascending transitions $\overline{R^{2}} = 0.987$ with $ SD=0.009 $ and for descending ones $ \overline{R^{2}}=0.986 $ with $ SD=0.016 $; see supplementary material for details. So we center our analysis on the quantity  $ f_{i+1}^2-f_{i}^2 $  instead $ f_{i+1}-f_{i} $, because this removes possible degenerations more efficiently, produces better fits to exponential functions, and generates larger relative distances between different values (see supplementary material for details).
In order to present ascending and descending transitions in the same histogram, we merge the left and right branches of the distribution functions by defining the bin width as the average of the two bin widths. Then the probability distribution of $ f_{i+1}^{2}-f_{i}^{2} $ can be written as
\begin{equation}\label{ProbHistograms}
P(\varepsilon)=\begin{cases}
F_{+}^{H}e^{-\varepsilon/ G_{+}^{H}} & \text{for $ \varepsilon > 0 $}\\F_{-}^{H}e^{\varepsilon/ G_{-}^{H}} & \text{for $ \varepsilon < 0 $}
\end{cases}\quad,  
\end{equation}
where the notation $ \varepsilon $ emphasizes that these distributions are constructed over bins. In the case of the cumulative distributions, we use  CCDF for ascending transitions and complementary cumulative distribution functions for descending ones (CDF). These choices allow conserving the same functional form as the probability distributions (this is possible due to the exponential behavior) but without the use of bins:
\begin{equation}\label{ProbCumulatives}
\begin{split}
\begin{gathered}
P(f_{i+1}^2-f_{i}^2)=\begin{cases}
F_{+}^{C}e^{-(f_{i+1}^2-f_{i}^2)}/G_{+}^{C}&\text{for $(f_{i+1}^2-f_{i}^2)>0$}\\F_{-}^{C}e^{(f_{i+1}^2-f_{i}^2)}/G_{-}^{C}&\text{for $ (f_{i+1}^2-f_{i}^2) < 0 $}
\end{cases}\quad,
\end{gathered}
\end{split}
\end{equation}
the ``.xlsx'' file in supplementary material contains the values of $F_{+}^{H}, F_{-}^{H}, G_{+}^{H}, G_{-}^{H}, F_{+}^{C}, F_{-}^{C}, G_{+}^{C}, G_{-}^{C}$ of fits and the determination coefficients $ R^{2} $. These probability distributions resemble the asymmetric Laplace probability distribution with different amplitudes for positive and negative branches generating a discontinuity in the origin (Figure \ref{AsymmetricLaplace}) \cite{Kotz}.
\\
Figure \ref*{TempScalePartita} shows the histogram of the probability distribution of ascending and descending transitions (including unisons) in the case of the first movement of the \textit{Partita in A minor BWV 1013} as well as the bin degeneration for the corresponding \textit{ambitus}. To explain the effect of bin degeneration notice that the distance in $ Hz^{2} $ between pairs of differences $ f_{j}^2-f_{i}^2 $ for the twelve-tone equal-tempered scale varies in such a way that the number of differences inside an arbitrary bin $ \varepsilon $, its degeneracy, decreases when $ |f_{j}^2-f_{i}^2| $  increases, and the probability distribution is equivalent to the corresponding one of a random melodic line (see supplementary material). The comparison between the distributions of real melodic lines with those from bin degeneration for the corresponding \textit{ambitus} indicates that the scale contributes to the observed results but does not explain them. Additionally, the probability distribution for bin degeneration fits better to a power law function ($ R^{2}=0.963$) than to an exponential function ($ R^{2}=0.934$) (see supplementary material for details).

The quantitative difference between the probability distribution for a real melodic line and its corresponding random one gives information about the order introduced by the composer coming from the selection of successive pairs of pitches. A mathematical tool for comparing two probability distributions is the Kullback-Leibler divergence or relative entropy \cite{Cover}
\begin{equation}\label{RelativeEntropy}
D_{KL}=\sum_{k=1}^{N}p_{k}\ln \left( \dfrac{p_{k}}{q_{k}} \right)\quad ,
\end{equation} 
where $ p_{k} $ is the probability distribution for the real melodic line to be compared with the \textit{a priori} distribution $ q_{k} $ coming from the degeneration of the $ k^{th} $ bin and that we associated with a random melodic line with the same \textit{ambitus} of the real one. $ N $ is the number of bins coming from the \textit{ambitus} with $ N/2 $ bins for each branch.    

\section{Statistical model}
\begin{figure*}[htb]
	\centering
	\begin{tabular}{c}
		\includegraphics[width=5.5cm]{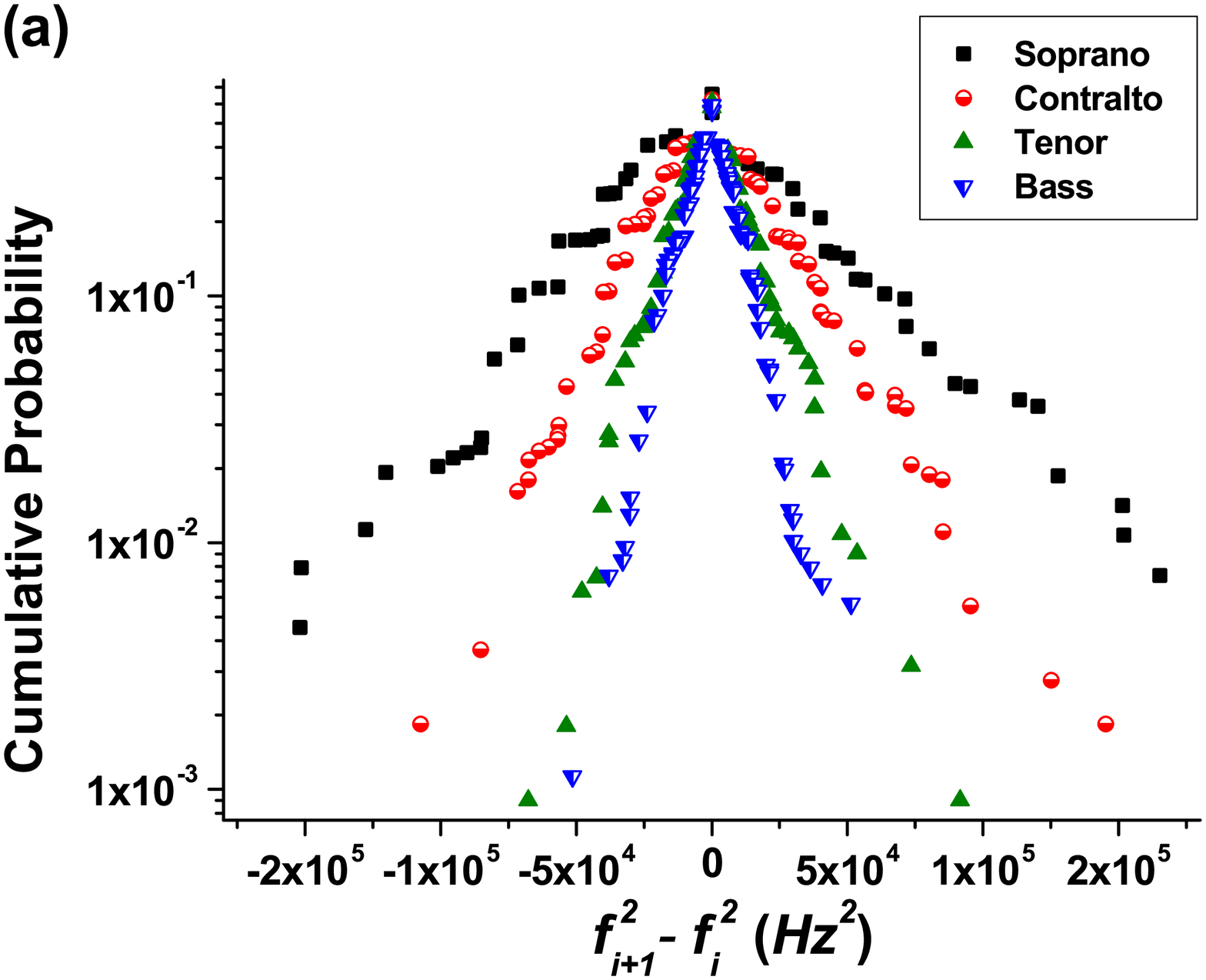}
	\includegraphics[width=5.5cm]{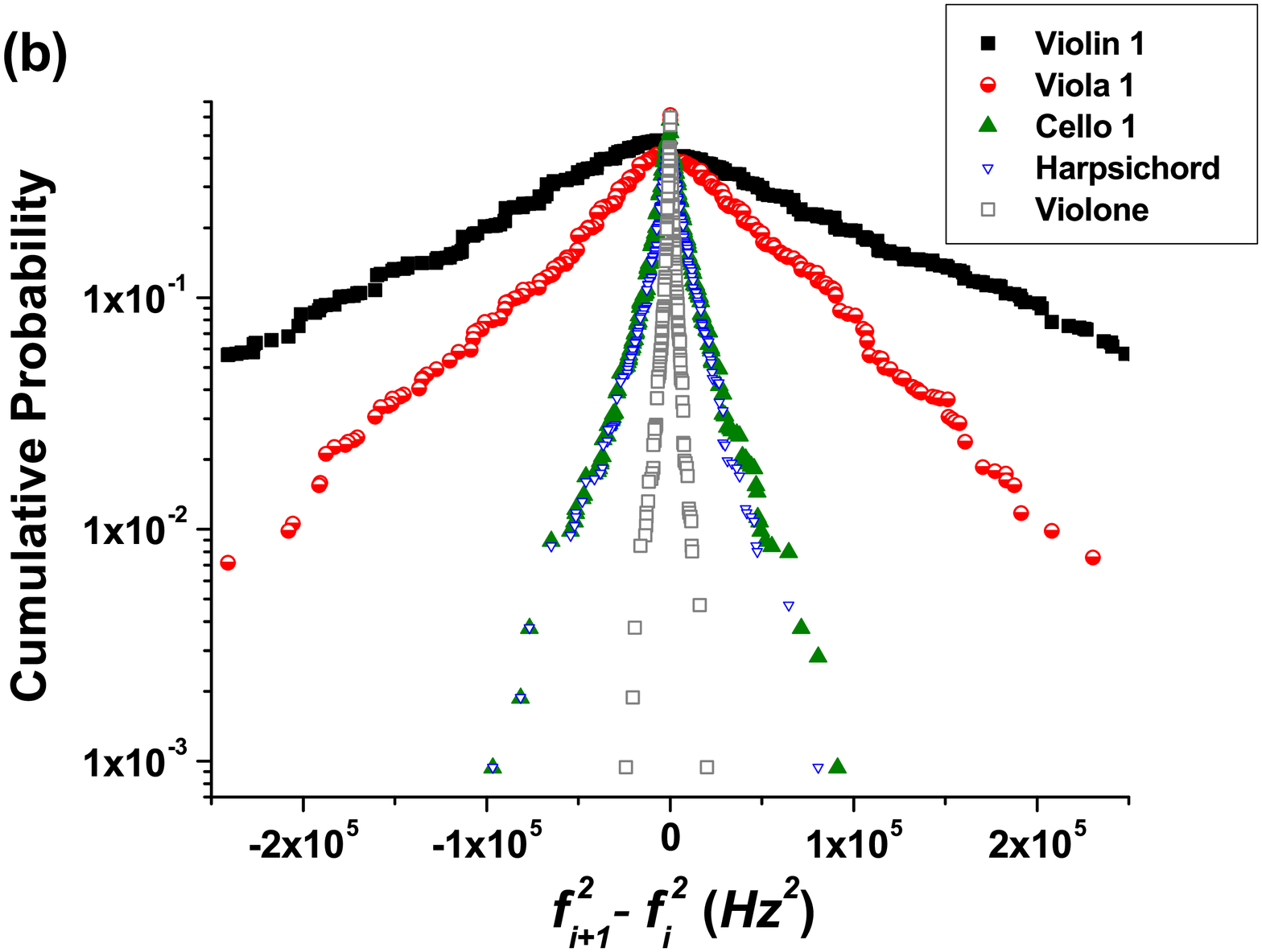}
\includegraphics[width=5.5cm]{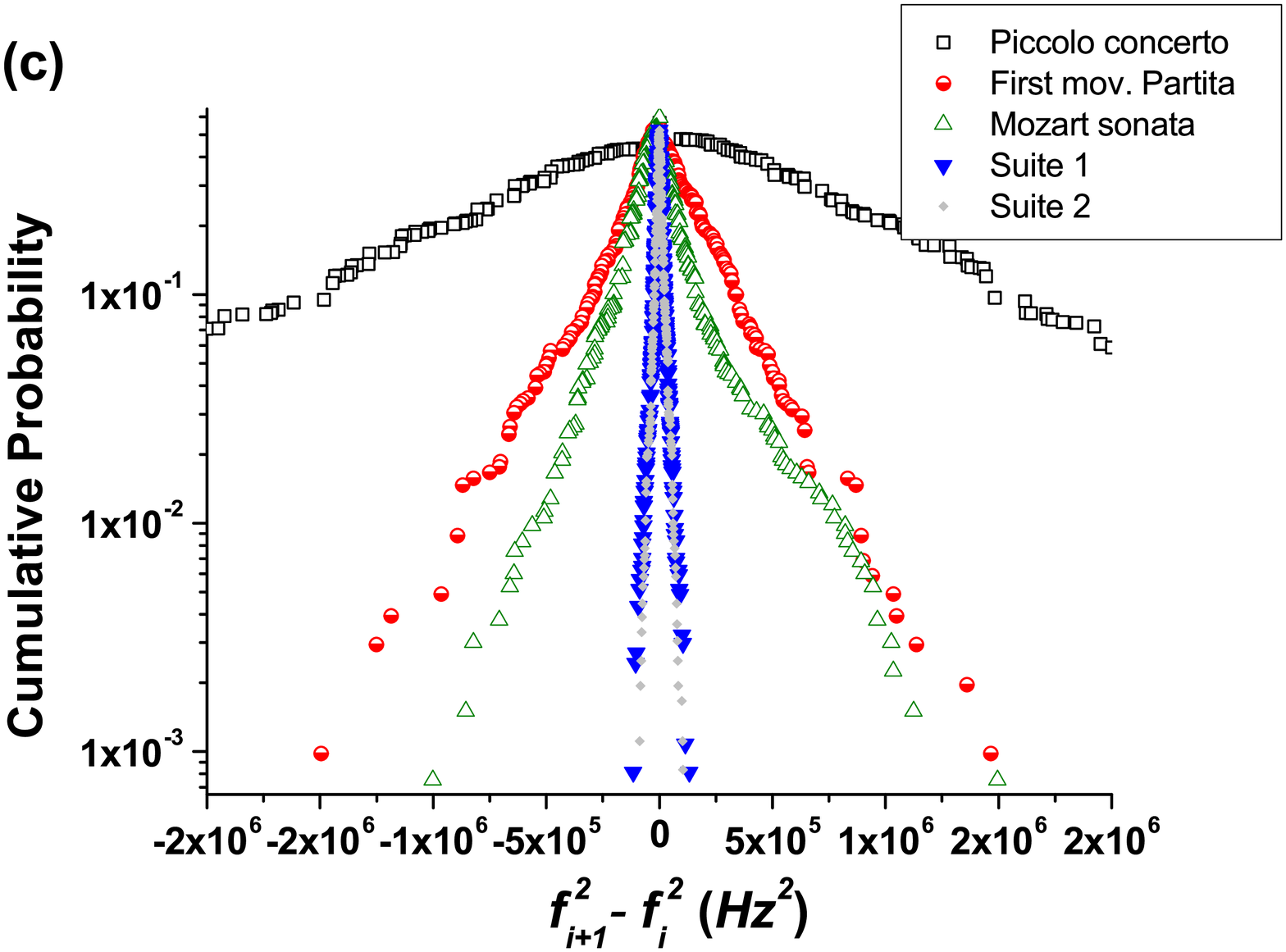}\\
\includegraphics[width=5.5cm]{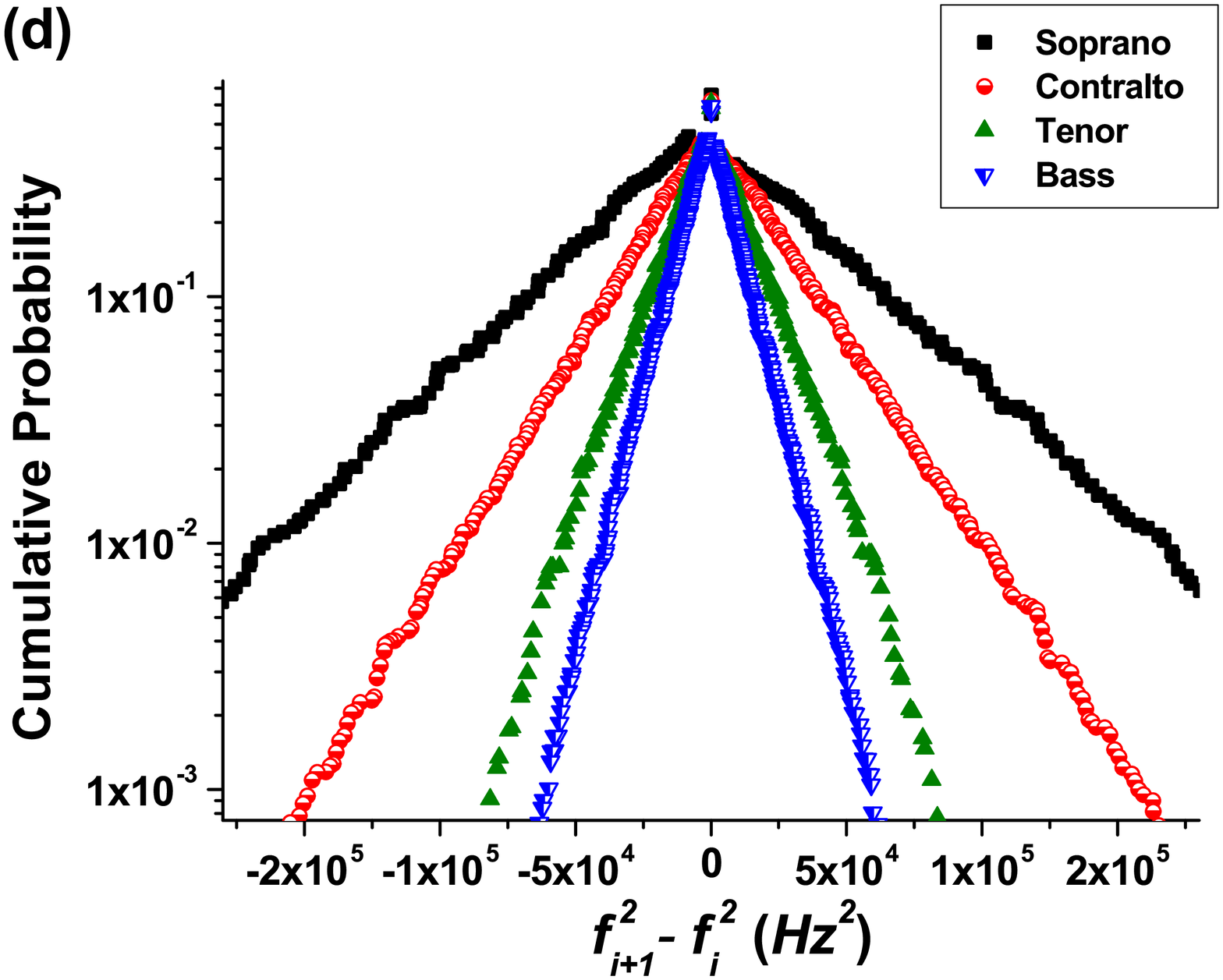}
\includegraphics[width=5.5cm]{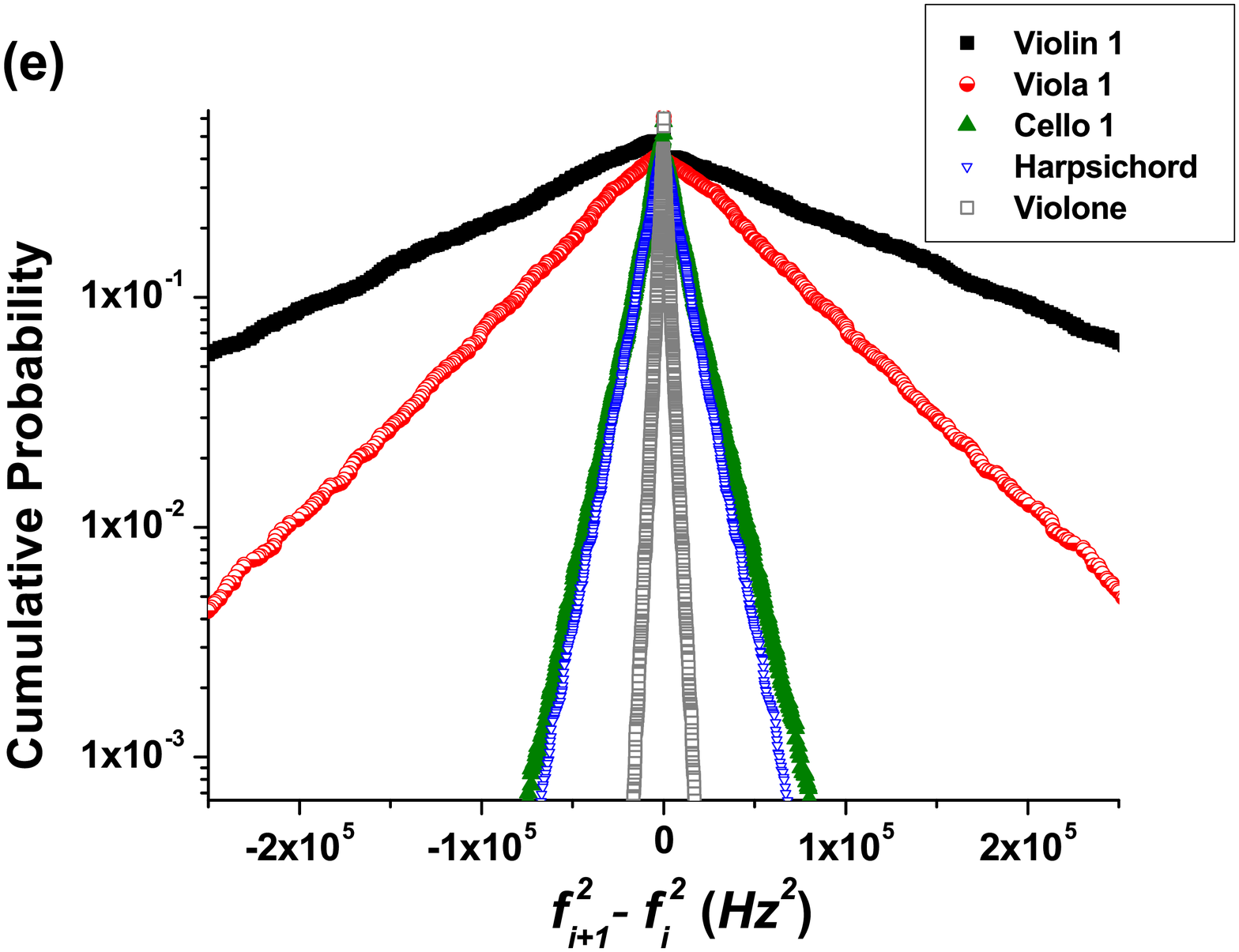}
\includegraphics[width=5.5cm]{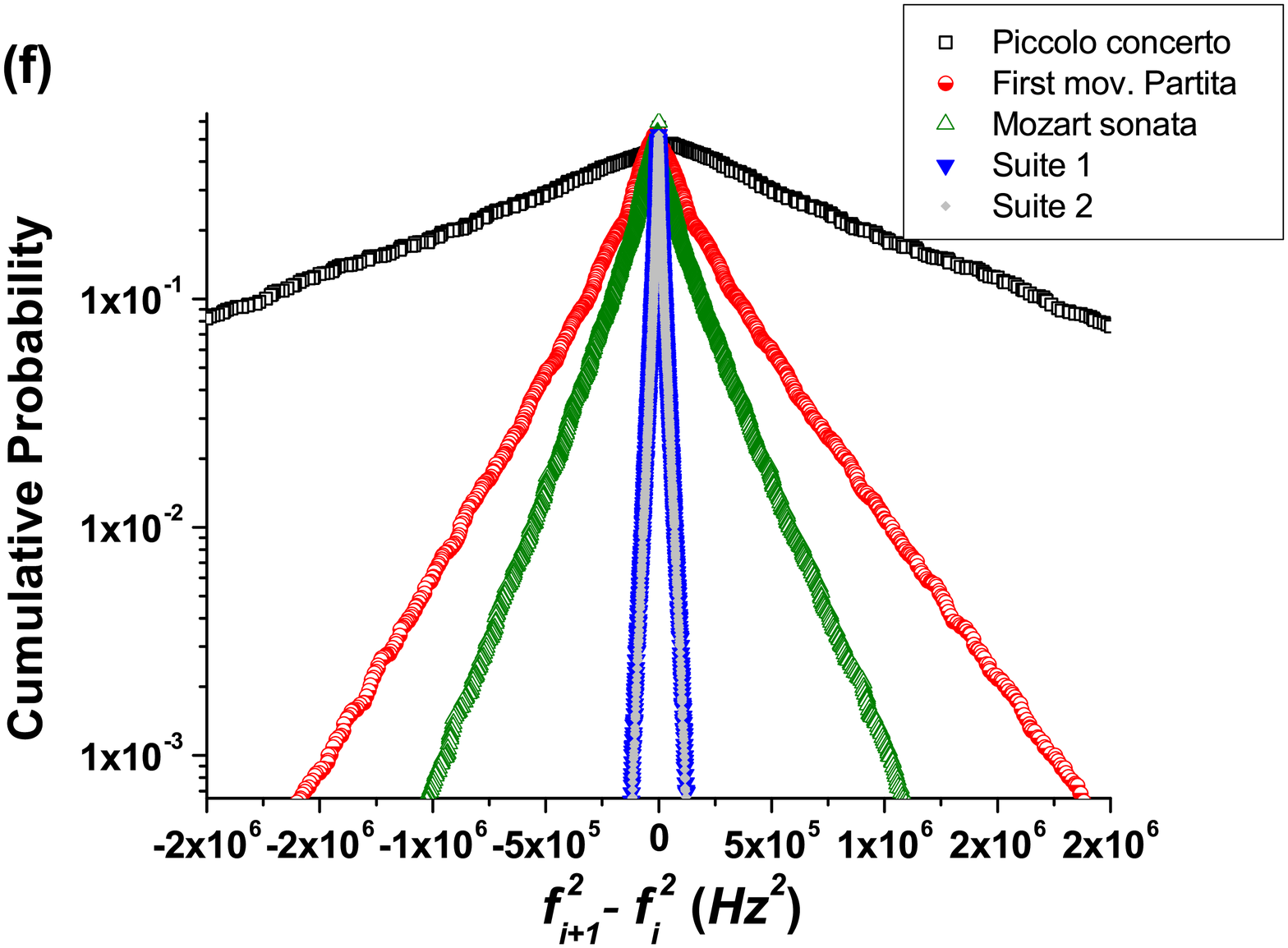}
\end{tabular}
\caption{Complementary cumulative distribution functions (ascending branches) and cumulative distribution functions (descending branches) for the empirical distributions (a, b, c) and the corresponding statistical model results (d, e, f). (a, d) \textit{Missa Super Dixit Maria} by Hans Leo Hassler. (b, e) \textit{Brandenburg Concerto No. 3 in G Major BWV 1048} by J. S. Bach, and (c, f) \textit{Piccolo Concerto RV444} by Antonio Vivaldi; \textit{First movement of the Partita in A Minor BWV 1013} by J. S. Bach; \textit{Sonata KV 545} by W. A. Mozart; \textit{Suite No. 1 in G Major BWV 1007} by J. S. Bach and \textit{Suite No. 2 in D Minor BWV 1008} by J. S. Bach. }\label{Results}  
\end{figure*}

For modeling the system in an analogy to equilibrium statistical physics, we use the definitions of melody \cite{Apel,Patel} and the results obtained by G. G\"und\"uz and U. G\"und\"uz \cite{Gunduz}, assuming that the composer would have to create a melodic line among the richest ones in terms of the combination of successive pitches. This procedure is equivalent to minimizing the relative entropy (the closest $ p_{k} $ to $q_{k}$) with mathematical constraints containing relevant musical information. 
The first constraint of the model comes from normalization. Since there are three kinds of different melodic intervals (ascending, descending, and unisons), then the normalization constraint is given by $ p_{a}+p_{d}+p_{u}=1 $, where $ p_{a} $ is the probability of ascending intervals, $ p_{d} $ is the probability of the descending ones and $ p_{u} $ the probability of unisons. If $ p_{u} $ is known, then $p_{a}$ and $p_{d}$ can be found from the normalization constraint and the asymmetry between ascending and descending intervals $ p_{a}-p_{d}  $. An equivalent alternative is to use the normalization constraint and to measure from histograms the next two observables
\begin{equation}\label{Constraints1}
 p_{d}+ p_{u}=\sum_{k=1}^{N/2}p_{k}\quad\text{and}\quad p_{a}+ p_{u}=\sum_{k=(\frac{N}{2}+1)}^{N}p_{k}\quad.
\end{equation} 
Here $ \sum_{k=1}^{N}p_{k}=1+p_{u} $ in order to be consistent with the double count of unisons. This choice only changes the result of the minimization problem (with respect to the normalized probability $ p^{*}_{k} = \tfrac{p_{k}}{1+p_{u}}$) by the multiplication of a positive constant (see supplementary material for details).  
\\The difference between the probability distributions of real melodic lines and their corresponding random ones leads to the second constraint. It comes from the selection made by the composer of the size and position in the register of musical intervals, independently of whether they are ascending or descending ones, which can be associated with the use of tonal consonance as follows. From Equation \ref{SumDifferenceL} with $ b\approx 1 $ and assuming that each bin $ k $ can be represented by the median of its extremes $ \varepsilon_{k} $ (in supplementary material we analyze this assumption for a typical case), since $ \varepsilon_{k} $ is linked with the interval size $ L $ and the tonal consonance parameter $ |f_{j}-f_{i} | $, then the expected value of $ |\varepsilon| $ can be written (see supplementary material for details) as:
\begin{equation}\label{Constraints2}
\begin{split}
\begin{gathered}
\langle|\varepsilon|\rangle=\sum_{k=1}^{N}p_{k}\cdot|\varepsilon_{k}|=a\Biggl[\sum_{k=1}^{N}p_{k}\langle\sigma^{2}/L\rangle_{k}+\sum_{k=1}^{N}p_{k}\langle\overline{|f_{j}-f_{i}|}^{2} / L\rangle_{k}\Biggl]\quad,
\end{gathered}
\end{split}
\end{equation}  
where $ \sigma^{2} $ is a variance associated with the size of intervals inside a bin. Using histograms, this is the best estimation that we can make of $ \langle|\varepsilon|\rangle $. The first term in the second line of Equation \ref{Constraints2} is related to the dispersion of the size of the musical interval inside each bin and the second term to the location of intervals in the register for each bin. The location of musical intervals is measured in terms of the average of the frequency differences, which can be related to an average of tonal consonance using the tonal consonance curves for complex tones (see supplementary material for the details). 
In this model, the expected value $ \langle|\varepsilon|\rangle $ is conserved, just as the expected value of the energy in the canonical ensemble of statistical physics \cite{Diogenes}. 
\\In order to explore the nature of the quantity $ \langle|\varepsilon|\rangle $, we consider the case in which each bin contains at maximum one possible transition. If we are in an equal amplitudes domain, then (see Equation \ref{Energy}) 
\begin{equation}\label{Constraint2Energy}
\langle|\varepsilon|\rangle=\langle|f_{p+1}^2-f_{p}^2|\rangle=|\epsilon_{p+1}^2-\epsilon_{p}^2|/(2\pi^{2}\rho T)\quad, 
\end{equation} 
where $ |\epsilon_{p+1}^2-\epsilon_{p}^2| $ will be the expected value of the magnitude of the difference in the average density of the total energy carried by the two fundamental components of the sound waves. Furthermore, the analogous expression of Equation \ref{Constraints2} 
when each bin contains a maximum of one possible transition will be:
\begin{equation}\label{Constraint2WithoutBins}
\begin{split}
\begin{gathered}
\langle|\varepsilon|\rangle=a\Biggl[\sum_{k=1}^{N}p_{k}\frac{(f_{j}-f_{i})_{k}^{2}}{L_{k}}\Biggl]
=a\Biggl[\sum_{x=1}^{W}\frac{1}{L_{x}}\sum_{h}p_{x_{h}}\frac{|f_{j}-f_{i}|^{2}}{x_{h}}\Biggl]\;\quad,
\end{gathered}
\end{split}
\end{equation}
where $ x $ runs from $ 1 $ to the total number of possible interval sizes $ W $ in the corresponding \textit{ambitus}, $ h $ refers to the possible locations of musical intervals of size $ x $ in the register, and $ p_{x_{h}} $ is the probability of finding a particular interval of size $ x $ in the position $ h $ of the register inside the melodic line. In comparison with Equation \ref{Constraints2}, the first term vanishes and the average disappears. Note that in Equation \ref{Constraint2WithoutBins} we could split the expected value into contributions of equal size $ L $, so we can use the tonal consonance curves for complex tones to directly relate the absolute value of frequency difference with the dissonance level. 
\\We include the asymmetry in the selection of ascending and descending intervals as a third constraint. This asymmetry is present in the difference of the coefficients for the left and right branches in Equations \ref{ProbHistograms} and \ref{ProbCumulatives}. First we analyze the continuity equation resulting from the fact that the second pitch of a transition is the first pitch of a successive transition. For a continuous segment of $ M $ successive pitches, the quantity $ \delta $ for this segment is defined as
\begin{equation}\label{Continuity}
\delta=f_{M}^{2}-f_{1}^{2}=\sum_{r=1}^{M-1}f_{r+1}^{2}-f_{r}^{2}\;\quad , 
\end{equation}
where $ r $ is a natural number that indicates the chronological order of appearance of pitches. The magnitude of $ \delta $ is smaller than or equal to the difference $ f_{max}^2-f_{min}^2 $, given by the maximum and the minimum frequencies in the \textit{ambitus} of the melodic line. If we have $ G $ transitions and $ U $ continuous segments separated by rests in a melodic line, the expected value of $ f_{r+1}^2-f_{r}^2 $ is
\begin{equation}\label{Constraint3WithoutBins}
\langle|f_{r+1}^2-f_{r}^2|\rangle=\frac{1}{G}\sum_{v=1}^{U}\delta_{v}\;\quad.
\end{equation}
Melodic balance (melodies tend to meander around a central pitch range) and the continuity of musical phrases are common properties of musical pieces, including those from the Baroque and Classical periods \cite{Aldwell}, leading to values of $ \langle|f_{r+1}^2-f_{r}^2|\rangle $
much smaller than those of $\langle f_{r+1}^2-f_{r}^2\rangle $. Using histograms, the best estimation that we can make of the expected value of Equation \ref{Constraint3WithoutBins} is
\begin{equation}\label{Constraints3}
\begin{split}
\begin{gathered}
\langle\varepsilon\rangle=\sum_{k=1}^{N}p_{k}\cdot\varepsilon_{k}=-a\Biggl[\sum_{k=1}^{N/2}p_{k}\langle\sigma^{2}/L\rangle_{k}+\sum_{k=1}^{N/2}p_{k}\langle\overline{|f_{j}-f_{i}|}^{2}/L\rangle_{k}\Biggl]
\\+a\Biggl[\sum_{k=(\frac{N}{2}+1)}^{N}p_{k}\langle\sigma^{2}/L\rangle_{k}+\sum_{k=(\frac{N}{2}+1)}^{N}p_{k}\langle\overline{|f_{j}-f_{i}|}^{2}/L\rangle_{k}\Biggl]\;\quad.
\end{gathered}
\end{split}
\end{equation}
Equation \ref{Constraints3} is our last constraint and shows that $ \langle\varepsilon\rangle  $ is due to the asymmetry in the use of the size and position in the register of ascending (positive terms) and descending (negative terms) intervals. Equations \ref{Constraints2} and \ref{Constraints3} imply that unless there is an extremely unusual asymmetry between the use of ascending and descending intervals, $ \langle\varepsilon\rangle $ must be much smaller than $\langle|\varepsilon|\rangle $, a behavior  that we noted in our experimental analysis (see ".xlsx" file in supplementary material).
Minimizing the relative entropy subject to the constraints in Equations \ref{Constraints1}, \ref{Constraints2}, and \ref{Constraints3} produced the probability distribution \cite{Kotz} (see supplementary material for details) 

\begin{equation}\label{LaplaceModel}
\begin{split}
\begin{gathered}
p_{k}=\begin{cases}
\dfrac{(p_{d}+p_{u})q_{k}e^{(-\lambda_{1}|\varepsilon_{k}|-\lambda_{2}\varepsilon_{k})}}{\sum\limits_{m=1}^{N/2}[q_{m}e^{(-\lambda_{1}|\varepsilon_{m}|-\lambda_{2}\varepsilon_{m})}]}&\text{for $k\in[1,N/2]$}\\ \\
\dfrac{(p_{a}+p_{u})q_{k}e^{(-\lambda_{1}|\varepsilon_{k}|-\lambda_{2}\varepsilon_{k})}}{\sum\limits_{m=(\frac{N}{2}+1)}^{N}[q_{m}e^{(-\lambda_{1}|\varepsilon_{m}|-\lambda_{2}\varepsilon_{m})}]}&\text{for $k\in[\frac{N}{2}+1,N]$},
\end{cases}
\end{gathered}
\end{split}
\end{equation}
where $ \lambda_{1} $ and $ \lambda_{2} $ are the Lagrange multipliers for constraints \ref{Constraints2} and \ref{Constraints3}, respectively. Using the expected values $ \langle|\varepsilon|\rangle  $ and $ \langle\varepsilon\rangle  $ obtained from the empirical distributions for the selected melodic lines, and allowing less than $ 1.0\% $ of error between the results from the statistical model and those of real data, we obtain the values for $ \lambda_{1} $ and $ \lambda_{2} $ (see ".xlsx" file in supplementary material), with $ \lambda_{1} $ from one to two orders of magnitude larger than $ \lambda_{2} $. While the values of $ \lambda_{1} $ are positive, those of $ \lambda_{2} $ can be positive or negative, showing possible asymmetries in the use of ascending and descending intervals. We obtained negative values of $ \lambda_{2} $ and $ p_{a}-p_{d} $ in almost all cases (the \textit{Piccolo Concerto RV444} of Antonio Vivaldi is the exception). These two behaviors generate, for the ascending and descending branches, different decay coefficients and different intercept points with the ordinate axis, in agreement with the asymmetry reported in many cultures in the sense that large melodic intervals are more likely to ascend and small melodic intervals are more likely to descend \cite{Huron} in the process of meandering around a central pitch range. Figure \ref{AsymmetricLaplace} was constructed with the purpose of representing these particular asymmetries in our model: $ P_{1}>P_{2} $ and $ \alpha_{1}>\alpha_{2}$ (implying that $\lambda_{2}<0$).     
\\With respect to the quantitative results of the model, each fit parameter is of the same order of magnitude for real melodic lines and those from the statistical model, and most of them are consistent within the error bars of the fits (see ".xlsx" file in supplementary material). 
Figure \ref{Suite2Bach} shows the comparison between the statistical model and the empirical results in the case of the Suite No. 2 BWV 1008. Some differences between the empirical data and the results from the statistical model are expected, since there are patterns in real melodic lines that cannot be captured by this model.
\begin{figure}[htb]
	\centering
	\begin{tabular}{c}
		\includegraphics[width=8.0cm]{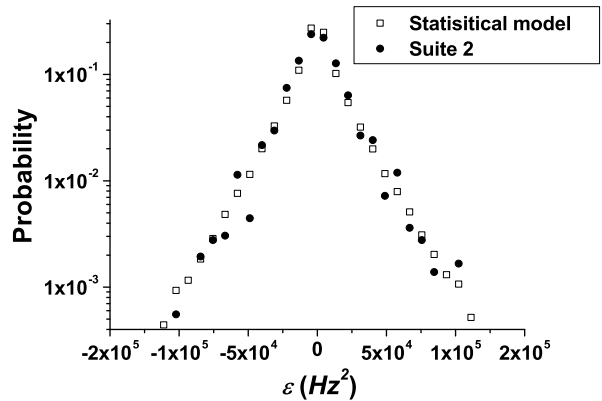}\end{tabular}
	\caption{Comparison in histograms between the  melodic line of \textit{Suite No. 2 BWV 1008} by Johann Sebastian Bach and that generated by the statistical model.}\label{Suite2Bach}  
\end{figure}
\\In order to compare probability distributions of different melodic lines with different bin widths in the histogram, we use the CCDF (ascending branch) and CDF (descending branch). For the statistical model, CCDF and CDF can be generated by dividing the probability of a bin between all the possible transitions inside it. The possible transitions are generated using the \textit{ambitus} of the corresponding melodic line, and we randomly distribute the probability inside each bin. Figure \ref{Results} shows the CCDF and CDF for the empirical data and the corresponding results from the statistical model for all studied melodic lines. In this figure, some features can be noticed; for example, different registers of musical instruments or human voices can be distinguished using the Lagrange multiplier $ \lambda_{1} $ to locate the approximate region of the exponential decay of both branches (without taking into account the asymmetry effect, which is weaker). This behavior allows distinguishing between the same melodic line played in different parts of the register, a musical process called transposition \cite{Apel}. In the \textit{Brandenburg Concerto No. 3 in G Major BWV 1048} by J. S. Bach, the harpsichord plays the same melodic line as the violone, but  transposed one octave higher. This change can be observed in the exponential decay parameters of the empirical distributions or the Lagrange multipliers in the statistical model (see ".xlsx" file in supplementary material). 
Additionally, since the Lagrange multiplier $ \lambda_{1} $ locates the approximate region of the exponential decay and each region is related to the tonal consonance properties of melodic intervals (compare Figure \ref{Results} with Figure \ref{ConsonanceCurve}b), it can be used as an indicator of the tonal consonance properties of a melodic line. Comparing Figures \ref{Results} and \ref{ConsonanceCurve}b, it is possible to understand why the same melody transposed to a lower part of the register is perceived as more dissonant \cite{Roderer}.

\section{Conclusions}
We found that the tonal consonance parameters for complex tones can be used to link the level of consonance perceived by humans with complexity in melody. Melodic lines are represented in terms of the physical parameters of tonal consonance and an entropy extremalization principle, with two macroscopic constraints that express the selection made by composers of melodic intervals and that can be interpreted in a musical sense: $\langle|\varepsilon|\rangle $ measures the average preference with respect to the size and the position in the register of melodic intervals, and $\langle\varepsilon\rangle $ measures the macroscopic asymmetry in the use of ascending and descending intervals. After the extremalization process subject to constraints, we obtained asymmetrical non-continuous Laplace distributions containing one Lagrange multiplier per constraint. One of the Lagrange multipliers ($ \lambda_{1} $) locates the region in which the melody is played,  capturing musical processes as transposition, registers of musical instruments (and human voices), and tonal consonance levels. The other Lagrange multiplier ($ \lambda_{2} $) captures asymmetry patterns between ascending and descending intervals reported for many cultures (and also found in our experimental study) in the sense that large melodic intervals are more likely to ascend and small melodic intervals are more likely to descend.
\\These findings show that some features of creativity in music can be modeled in an analogy to physical systems, using macroscopic rules similar to energy conservation, and principles such as entropy maximization. While many non-physical complex systems exhibit emergent properties, here the variable used for describing the microscopic properties of the system is a physical quantity that in the equal amplitude domain is functionally similar to the difference in the energy density carried by the fundamental components of two successive sound waves.\\
\section{Acknowledgments}
Universidad Nacional de Colombia funded this research under grant HERMES 19010. We thank Charles Barrett, Jack Crimmins and Daniel Rasolt for their comments on this manuscript.

\bibliography{Refs}
\bibliographystyle{unsrt}

\end{document}